\documentclass[fleqn,12pt,twoside]{article}
\usepackage{espcrc1}

\usepackage{graphicx}
\usepackage[figuresright]{rotating}

\newcommand{\AmS}{{\protect\the\textfont2
  A\kern-.1667em\lower.5ex\hbox{M}\kern-.125emS}}

\title{Measurement of the $Q^2$-evolution of the Bjorken integral  
and extraction of an effective strong coupling constant at low $Q^2$.}

\author{A. Deur
\address{Thomas Jefferson National Accelerator Facility, USA}}
       
\begin{document}

\maketitle

\begin{abstract}
We report on the measurement of the Bjorken sum in the range $0.16<Q^2<1.1$ GeV$^2$.
The extraction of an effective strong coupling constant is then discussed.
\end{abstract}

\section{Bjorken Sum Rule}

The Bjorken sum rule \cite{Bjorken} has been of central importance  
for studying the spin structure of the nucleon. Accounting for
finite $Q^2$ corrections to the sum rule, it reads:
\begin{equation}
\vspace*{-0.2cm}
\int_{0}^{1}(g_{1}^{p}-g_{1}^{n})dx=\frac{g_{a}}{6}
[ 1-\frac{\alpha_{\rm{s}}}{\pi}-3.58\left(\frac{\alpha_{\rm{s}}}
{\pi}\right)^{2}-20.21\left(\frac{\alpha_{\rm{s}}}{\pi}\right)^{3} +...]
+\sum_{i=1}^\infty {\mu^{p-n}_{2i+2} \over Q^{2i}},
\label{eqn:bj}
\end{equation}
\noindent 
where the $\mu^{p-n}_{2i+2}$ are higher twist terms. The sum rule has 
been checked experimentally at $Q^2$=5 GeV$^2$ to better than 10\%.
As recently realized, the Bjorken sum rule is related to a 
more general sum rule, the generalized Gerasimov-Drell-Hearn (GDH) sum 
rule \cite{GDH,gGDH}:
\begin{equation}
\vspace*{-0.2cm}
\int_{0}^{1^{-}}(g_{1}^{p}-g_{1}^{n})dx=
\frac{Q^2}{16\pi^2\alpha}(\rm{GDH}^p(Q^2)-\rm{GDH}^n(Q^2)).
\label{eqn:link}
\end{equation}
\noindent
Since the generalized GDH sum is, in principle, calculable 
at any $Q^2$, it can be used to study the transition from
the partonic to hadronic degrees of freedom of the strong force.  
However, the validity domains for chiral perturbation theory ($\chi$PT) 
at low $Q^2$ and pQCD calculations at higher $Q^2$ used to calculate the GDH 
sum do not overlap. The Bjorken sum  is the flavor non-singlet part 
of the GDH sum. This leads to simplifications that may help in linking the 
($\chi$PT) validity domain to the pQCD validity domain \cite{Volker}. 
Hence the Bjorken sum 
appears as a key quantity to study the hadron-parton transition. 

We used data from the Thomas Jefferson National Accelerator Facility 
(JLab) \cite{eg1a proton,eg1a deuteron,E94010} to extract the Bjorken sum from 
$Q^2=0.16$ to 1.1 GeV$^2$ \cite{bj} (Fig. 1 panel A). At low $Q^2$, 
we can compare to $\chi$PT calculations \cite{jichipt,meissnerchipt}. The data 
agree well with models \cite{ST,AO} and also with Eq.~1 calculated to third 
order in $\alpha_s$ and to leading twist. The agreement 
between the data and the leading twist calculation down to quite low $Q^2$ 
indicates that overall higher twist effects \cite{bj} are small. 

\section{ The Effective Strong Coupling Constant, $\alpha_{\rm{s}}^{\rm{eff}}$}
\vspace*{-0.2cm}
$\alpha_{\rm{s}}$ can be extracted from Eq.~1 if higher
twists are known or negligible. This is not the case here. This difficulty 
disappears when considering effective coupling constants \cite{grunberg}. 
In that case, $\alpha_{\rm{s}}^{\rm{eff}}$ contains higher twists and QCD 
radiation effects. As a consequence, $\alpha_{\rm{s}}^{\rm{eff}}$ is
analytical at any $Q^2$ and renormalization 
scheme independent. However, $\alpha_{\rm{s}}^{\rm{eff}}$ becomes process 
dependent which is not a problem since these different coupling constants are 
related by ``commensurate scale relations'' that connect observables without 
scheme or scale ambiguity \cite{brodsky1,brodsky11}. Following this procedure,
$\alpha_{\rm{s}}^{\rm{eff}}$ is extracted using the equation:
\vspace{-0.2cm}
\begin{eqnarray}
\Gamma_{1}^{p-n}=\frac{1}{6}g_{a} [ 1-\frac{\alpha^{eff}_{s}}{\pi} ].
\label{eqn:bjltlo}
\end{eqnarray}
\noindent 
Such $\alpha_{\rm{s}}^{\rm{eff}}$ is shown in Fig.~1B, together with 
$\alpha_{\rm{s}}^{\rm{eff}}$ extracted
using $\Gamma_{1}^{p-n}$ from Eq.~1 computed to third order in 
$\alpha_s$ and to leading twist. Also shown are $\alpha_{\rm{s}}$ calculated 
to order $\beta_0$,
$\alpha_{\rm{s}}^{\rm{eff}}$ calculated with the model \cite{AO}, 
and $\alpha_{\rm{s}}^{\rm{eff}}$ extracted from world data on the Bjorken
sum. 
$\alpha_{\rm{s}}^{\rm{eff}}$ merges with $\alpha_{\rm{s}}$ at large $Q^2$
since their difference is due to higher twists and gluon bremsstrahlung. At 
$Q^2 = 0$, the GDH sum rule gives the slope of 
$\alpha_{\rm{s}}^{\rm{eff}}$. 
The data, together with the constraint at $Q^2 \simeq 0$, 
hint that $\alpha_{\rm{s}}^{\rm{eff}}$ has no $Q^2$ scale 
dependence at low $Q^2$.

In QED or QCD, only loops on the exchanged photon or gluon are responsible 
for the running of the coupling constant because of the Ward identities. 
Consequently, theoretical calculations of the running coupling constant deal
only with dressed propagators. We can assume that, in order to compare to 
non-perturbative calculations of $\alpha_{\rm{s}}^{\rm{eff}}$, we do not need 
to include in $\alpha_{\rm{s}}^{\rm{eff}}$ the gluon bremsstrahlung and
vertex corrections. This amounts to not folding the QCD radiative corrections 
into $\alpha_{\rm{s}}^{\rm{eff}}$ and redefining it using the equation:

\vspace{-0.5cm}
\begin{eqnarray}
\Gamma_{1}^{p-n}=\frac{g_{a}}{6}[ 1-\frac{\alpha_{s}^{eff'}}{\pi}-3.58
\left(\frac{\alpha_{s}^{eff'}}{\pi}\right)^{2}
-20.21\left(\frac{\alpha_{s}^{eff'}}{\pi}\right)^{3}-130\left(\frac{
\alpha_{s}^{eff'}}{\pi}\right)^{4}
-893\left(\frac{\alpha_{s}^{eff'}}{\pi}\right)^{5} ]. \nonumber
\end{eqnarray}

\begin{figure}[ht!]
\begin{center}
\includegraphics[scale=0.4]{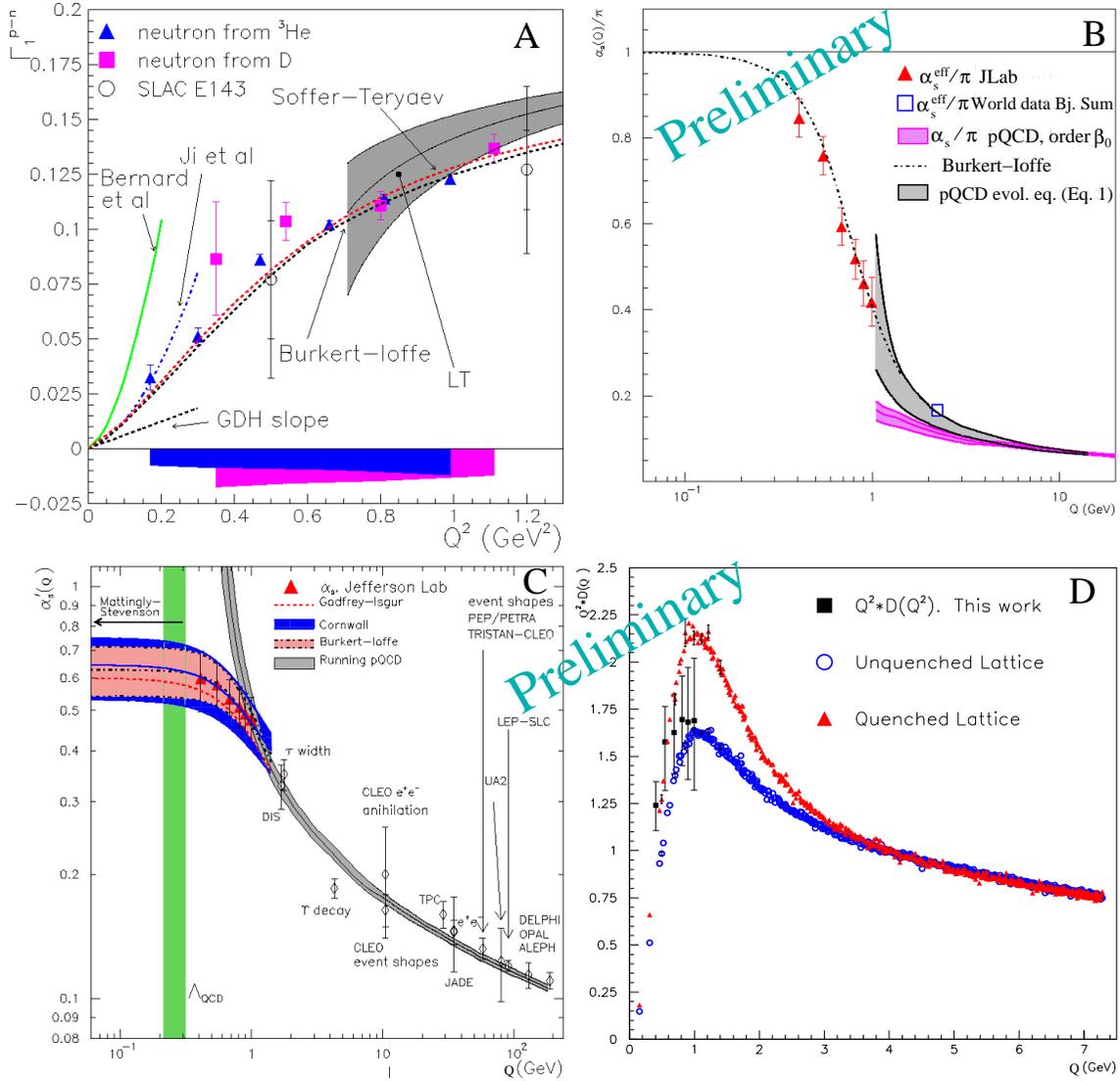}
\end{center}
\caption{Left top (Panel A): $Q^2$-evolution of the Bjorken sum. The dark (light)
horizontal band is the experimental systematic error corresponding to the
neutron extracted using $^3$He (D). Right top (Panel B): Effective 
strong coupling constant as defined by Eq.~3.
Left bottom (Panel C): Extracted $\alpha_{s}^{\rm{eff'}}(Q)$ together with 
experimental $\alpha_{s}(Q)$, running of pQCD and phenomenological 
$\alpha_{s}(Q)$.
The vertical band represents $\Lambda_{QCD}$ and its uncertainty. The dark
band gives the uncertainty on the Cornwall calculation due to $\Lambda_{QCD}$. 
The lighter band is the uncertainty on the Burkert-Ioffe model due to the 
truncation of the leading twist series to $5^{th}$ order. 
Right bottom (Panel D): The gluon transverse propagator multiplied by $Q^{2}$, 
extracted from our result together with quenched and unquenched Lattice QCD
calculations \cite{Bowman}.}

\label{fig:alglu}
\end{figure}

The error from the series truncation is estimated
by taking the difference between the fourth and fifth orders. With this
redefinition, $\alpha_{\rm{s}}^{\rm{eff'}}$ becomes scheme-dependent 
(we work in the $\overline{\textrm{MS}}$ 
scheme). The result is shown in the panel C of
Fig.~\ref{fig:alglu} along with world data, the running of $\alpha_{s}$ 
from pQCD and estimates of the phenomenological running constant. 
In ref.  \cite{Cornwall} a solution to the Dyson-Schwinger equations 
regularizes the infrared behavior of $\alpha_{s}$ by generating an effective 
gluon mass that is found to be $m_g=500\pm200$ MeV. For us, $m_g$ is 
constrained by the GDH sum rule which determines the derivative of the Bjorken 
integral at $Q^{2}=0$. This imposes 
$\alpha_{s}^{eff'}(Q^{2}=0)=0.629 \pm 0.086$ which in turn constrains the 
gluon mass at the photon point to be $407 \pm 51$ MeV. Mattingly and 
Stevenson \cite{Mattingly} 
used e$^{+}$/e$^{-}$ annihilation to extract an effective $\alpha_{s}$. 
The curve from Godfrey and Isgur \cite{Godfrey-Isgur} shows the coupling 
constant needed in their quark model to reproduce hadron spectroscopy.

Lattice QCD results present more often the gluon propagator rather than
the coupling constant. Since the behavior of $\alpha_{s}^{eff'}$ at low 
$Q^2$ may be accounted for by a dynamical gluon mass, 
which modifies the gluon propagator, we can extract from our result 
an ``effective gluon propagator'' and compare it to Lattice calculations. 
The Dyson-Schwinger equations provide the non-perturbative approach needed 
for studying the gluon propagator. However, the necessity of approximations 
to solve the equations results in some ambiguity. We use the results of 
Cornwall \cite{Cornwall} which provide good comparison with results from 
various studies. Results on the gluon propagator multiplied by $Q^{2}$ are 
shown on the
panel D of Fig. \ref{fig:alglu}, along with quenched and unquenched Lattice 
QCD results \cite{Bowman}. More Lattice results exist but they
are mostly quenched and agree with Ref.  \cite{Bowman}. 

\section{Summary and Conclusion}
We have presented the Bjorken sum in the $Q^2$ range of 0.16-1.1 GeV$^2$. The 
gap between the parton to hadron descriptions of the strong interaction, if 
smaller, is not bridged yet.
With these data, we extracted an effective coupling for the strong interaction.
We hypothesize that $\alpha_{\rm{s}}^{\rm{eff'}}$ defined when QCD radiations
are not folded in can be compared to the various effective couplings available
from theories. These physical couplings,  
obtained within very different areas of strong interaction (hadron 
spectroscopy, non-perturbative calculations, lattice QCD and moments of 
structure functions) agree with our data. $\alpha_{\rm{s}}^{\rm{eff}}$ can be 
used to parametrize the strong force at any $Q^2$. Our data and the $Q^2=0$ 
constraint hint that $\alpha_{\rm{s}}^{\rm{eff}}$ loses its scale
dependence at very low $Q^2$. This will be checked by upcoming 
experimental results at very low $Q^2$  \cite{gdh neutron,gdh proton}.  
\vspace{-0.5cm}
\section{Acknowledgments}
This work was supported by the U.S. Department of Energy (DOE) and the U.S.
National Science Foundation. The Southeastern Universities Research 
Association operates the Thomas Jefferson National Accelerator 
Facility for the DOE under contract DE-AC05-84ER40150.

\vspace*{-1.cm}

\end{document}